\documentclass[prl,showpacs,aps,twocolumn]{revtex4-1}

\usepackage{amsmath, amssymb}
\usepackage{graphicx}

\newcommand{\tr}{\mathop{\mathrm{tr}}\nolimits}
\renewcommand{\section}[1]{{\par\it #1.---}\ignorespaces}
\begin{document}
\title{Steady-State Coherent Transfer by Adiabatic Passage}

\author{Jan Huneke}
\author{Gloria Platero}
\author{Sigmund Kohler}
\affiliation{Instituto de Ciencia de Materiales, CSIC,
	Cantoblanco, E-28049 Madrid, Spain}

\date{\today}


\begin{abstract}

We propose steady-state electron transport based on coherent transfer
by adiabatic passage (CTAP) in a linearly arranged triple quantum dot
with leads attached to the outer dots.  Its main feature is repeated
steering of single electrons from the first dot to the last dot
without relevant occupation of the middle dot.  The coupling to
leads enables a steady-state current, whose shot noise
is significantly suppressed provided that the CTAP protocol performs
properly.  This represents an indication for the direct transfer
between spatially separated dots and, thus, may resolve the problem of
finding experimental evidence for the non-occupation of the middle
dot.

\end{abstract}

\maketitle

The possibility of finding particles in delocalized states is a most
intriguing feature of quantum systems.  It gives rise to phenomena
such as tunneling which is intrinsically quantum mechanical and beyond
everyday experience.  A particular tunnel effect is the coherent
transfer by adiabatic passage (CTAP) \cite{Greentree2004a} of an
electron in a linearly arranged triple quantum dot
\cite{Gaudreau2006a, Schroer2007a}.  It is based on the control
of the inter-dot tunnelings, such that an electron is guided from the
first dot to the last dot without populating the middle dot.  While
being closely related to stimulated Raman adiabatic passage
\cite{Oreg1984a, Kuklinski1989a}, CTAP possesses the appealing feature
of dealing with spatially separated states.

The non-occupation of the middle dot can be theoretically predicted
straightforwardly by solving the Schr{\"o}dinger equation or, in the
presence of decoherence, an appropriate quantum master equation.  By
contrast, its direct experimental proof is less obvious, because
such position measurement unavoidably acts back on the position.
Thus, such measurement would influence the effect that it should
substantiate. For example, coupling the middle dot to a charge
detector \cite{Gustavsson2006a, Fricke2007a} impacts significantly on
the dot populations \cite{Rech2011a}, since it causes decoherence to
which CTAP is quite sensitive \cite{Kamleitner2008a}.  Consequently,
as a matter of principle, direct observation of the dot populations is
of limited use and, thus, one may have to resort to an indirect
verification.

\begin{figure}[b]
\centerline{\includegraphics{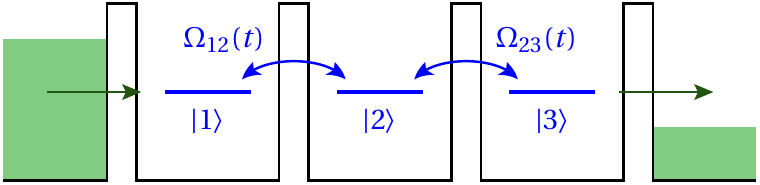}}
\caption{Triple quantum dot in linear arrangement, where the first and
the last dot are tunnel coupled to electron reservoirs with a large
voltage bias applied.  The CTAP protocol can be implemented by proper
time-dependent gate voltages at the inter-dot barriers, such that the
corresponding tunnel matrix elements $\Omega_{12}(t)$ and
$\Omega_{23}(t)$ obey Eq.~\eqref{pulse}.
}
\label{fig:setup}
\end{figure}

In this work, we propose a generalization of CTAP that circumvents
this difficulty and predicts an experimental fingerprint for the
non-occupation of the middle dot.  It is based on a modified setup
with electron reservoirs attached to the first and the last quantum
dot (see Fig.~\ref{fig:setup}), such that a current may flow.  We
demonstrate that the suggested steady-state CTAP protocol establishes
a rather regular electron transport with suppressed shot noise
indicating its proper course.  In turn, if the middle dot is
significantly occupied, transport becomes more noisy.
Our analysis of the transport process is based on a numerical
propagation of a time-dependent quantum master equation, which needs
to be augmented in order to compute current noise \cite{Kaiser2007a,
Sanchez2008c}.

\section{Model and master equation}
\begin{figure}
\centerline{\includegraphics{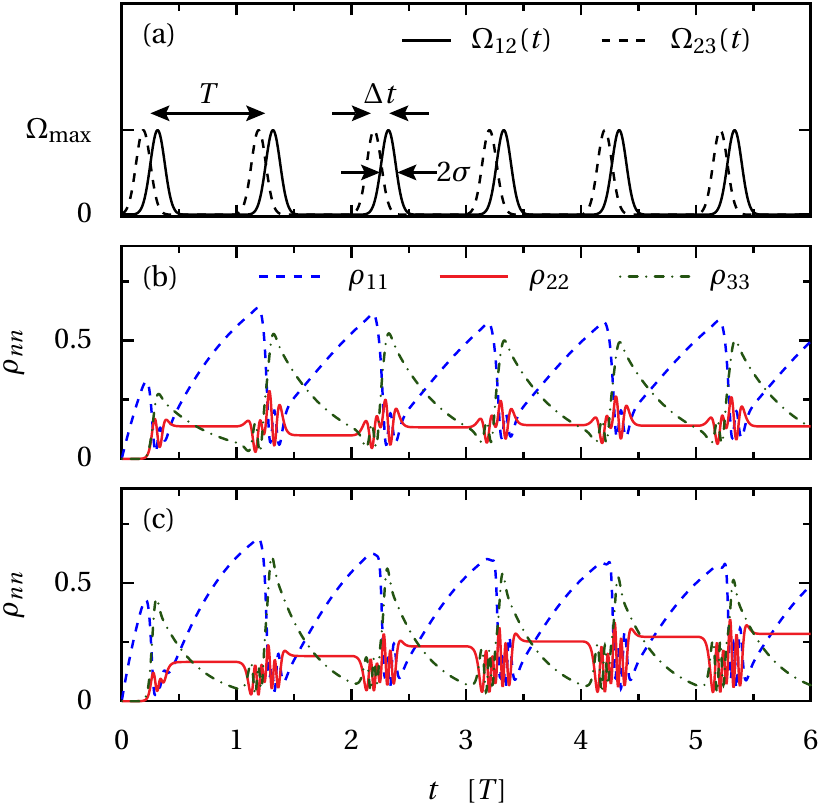}}
\caption{(Color online) (a) Time profile of the inter-dot tunnel
matrix elements forming repeated CTAP pulses.  (b,c) Time evolution of
the dot occupation probabilities for $\Gamma=0.5/\Omega_\text{max}$ and
driving periods (b) $T=2/\Omega_\text{max}$ and (c)
$T=2.9/\Omega_\text{max}$.}
\label{fig:timeevolution}
\end{figure}

The setup sketched in Fig.~\ref{fig:setup} is described by the
Hamiltonian $H(t) = H_\text{TQD}(t) + H_\text{dot-leads} +
H_L + H_R$, where (in units with $\hbar=1$)
\begin{equation}
\label{HTQD}
H_\text{TQD}(t)
= \Omega_{12}(t) c_1^\dagger c_2 + \Omega_{23}(t) c_2^\dagger c_3
+\text{H.c.}
\end{equation}
refers to the triple quantum dot with vanishing onsite energies.
We assume strong Coulomb repulsion such that the triple dot can be
occupied with at most one electron.  Hence the electron spin has
no relevant influence and will be ignored.  By applying appropriate gate
voltages to the inter-dot barriers, the tunnel matrix elements between
neighboring dots, $\Omega_{12}(t)$ and $\Omega_{23}(t)$, are
controlled such that they assume the sequence of double Gauss pulses
\begin{equation}
\label{pulse}
\Omega_{12/23}(t) = \sum_{n=0}^\infty
\Omega_\text{max} \exp\Big[-\frac{(t\pm\Delta t/2-nT-t_0)^2}{2\sigma^2}\Big] ,
\end{equation}
with amplitude $\Omega_\text{max}$, width $2\sigma$, delay time $\Delta
t$, and repetition time $T$ as is depicted in
Fig.~\ref{fig:timeevolution}(a).  This implies the time periodicity
$H(t) = H(t+T)$ for $t>t_0$ with some time-offest $t_0$.

Ideally, one would like to modify the control parameters $\Omega_{ij}$
adiabatically slowly to ensure ideal accomplishment of the protocol.
This however is not possible, because a lower limit to the CTAP
duration is set by decoherence, which may stem from substrate phonons,
a detector, or as in the present case the coupling to leads.  For the
transient dynamics during a total propagation time $t_\text{max}$
considered in Ref.~\cite{Greentree2004a}, decoherence is tolerable for
the pulse parameters $\Delta t = 2\sigma = T/4$.  Here we use
$T=t_\text{max}$ as pulse repetition time and focus on the steady
state.
An important feature of the CTAP double pulse is its
``counter-intuitive'' order \cite{Greentree2004a}, which means that
the coupling $\Omega_{23}$ is active before $\Omega_{12}$.  This
requirement can be understood upon noticing that $(\Omega_{23}, 0,
-\Omega_{12})^T$ is an eigenstate of the single-particle Hamiltonian
in the basis $\{|1\rangle, |2\rangle, |3\rangle\}$.  Between the
maxima of the double peak, this eigenstate changes from $(1,0,0)^T$ to
$(0,0,-1)^T$.  Thus, the adiabatic time evolution of an electron
starting in dot~1 is as deserved, namely it undergoes a direct
transition to dot~3 without populating dot~2.

The leads are modeled as free electron gases with the Hamiltonian
$H_\text{leads} = \sum_{\ell,q} \epsilon_q c_{\ell q}^\dagger c_{\ell
q}$, where $c_{\ell q}^\dagger$ creates an electron in state $q$ of
lead $\ell=L,R$.  The coupling of dot~1 to the left lead is given by
the tunnel Hamiltonian $H_{L} = \sum_{q} V_{Lq} c_{Lq}^\dagger c_1
+\mathrm{H.c.}$ and $H_{R}$ accordingly.  The lead coupling is fully
determined by its spectral density $\Gamma_L(\epsilon) = 2\pi\sum_q
|V_{Lq}|^2 \delta(\epsilon-\epsilon_q)$, which we assume within a
wide-band limit energy independent and equal for both leads,
$\Gamma_L(\epsilon) = \Gamma_R(\epsilon) \equiv \Gamma$.

If all relevant dot energies lie within the voltage window of the two leads,
electron transport becomes unidirectional.  Then one can derive for
the reduced triple dot density operator $\rho$ within a Bloch-Redfield
approach the master equation \cite{Gurvitz1996a}
\begin{equation}
\label{ME}
\dot\rho
= -i[H_\text{TQD}(t),\rho]
  +\Gamma\mathcal{D}(c_1^\dagger)\rho
  +\Gamma\mathcal{D}(c_3)\rho
\equiv \mathcal{L}(t)\rho
\end{equation}
with the Lindblad form $\mathcal{D}(x)\rho\equiv x\rho x^\dagger
-\frac{1}{2}x^\dagger x\rho - \frac12 x^\dagger x$.
The dissipative superoperators $\mathcal{D}(c_1^\dagger)$ and
$\mathcal{D}(c_3)$ describe incoherent tunneling from the left
lead (source) to dot~1 and from dot~3 to the right lead (drain),
respectively.

\section{Computing current noise by numerical propagation}
The fluctuations of the current can be characterized by the current
auto correlation function $S(\omega)$ which is typically measured at
low frequencies.  Therefore, we focus on the zero-frequency limit in
which the correlation function relates to the variance of the
transported charge according to $S \equiv S(0) = e^2 \lim_{t\to\infty}
(d/dt)\langle\Delta N^2(t)\rangle$ \cite{MacDonald1949a,
Blanter2000a}, where $e$ is the elementary
charge.  The variance can be computed with the help of a density
operator $R(\chi,t)$ that is augmented by counting variable $\chi$
and possesses the limit $R(\chi{\to}0,t)=\rho(t)$.  Its trace $\tr
R(\chi,t) = \langle e^{i\chi N}\rangle$ is the moment generating
function for the number of transported electrons $N$, i.e., $\langle
N^k\rangle = (\partial/\partial i\chi)^k \tr R|_{\chi=0}$.  For
unidirectional transport, $R$ obeys the master equation $\dot
R(\chi,t) = \mathcal{L}R(\chi,t) + (e^{i\chi}-1)\mathcal{J}R(\chi,t)$
\cite{Flindt2004a}, where $\mathcal{J}R = \Gamma c_3 R
c_3^\dagger$ is the jump operator contained in the Lindblad form.
Exploiting the $2\pi$ periodicity in $\chi$, one frequently continues
with a Fourier transformation which yields the so-called $N$-resolved
master equation \cite{Gurvitz1996a}.  Then however, a major technical
problem arises, because one has to consider many additional degrees of
freedom, such that a numerical solution becomes rather expensive.  A
more economic approach has been developed in Refs.~\cite{Kaiser2007a,
Sanchez2008c}.  By Taylor expansion up to second order, $R(\chi) =
\rho + i\chi R_1 -\frac{1}{2}\chi^2 R_2$, one obtains the equations of
motion $\dot R_1 = \mathcal{L}R_1 + \mathcal{J}\rho$ and $\dot R_2 =
\mathcal{L}R_2 +2\mathcal{J}R_1 +\mathcal{J}\rho$, which we solve by
numerical propagation.  Then we use the fact that $\tr
R(\chi)$ is the moment generating function, so that $(d/dt)\langle
N^k\rangle = \tr\dot R_k$, by which we finally obtain the current
$I = e\lim_{t\to\infty}\tr\dot R_1$ and the zero-frequency noise $S
= eI + \lim_{t\to\infty} 2e\tr(e\mathcal{J}R_1-IR_1)$.  For transport
in periodically time-dependent conductors, these quantities must be
considered in the average over the driving period \cite{Kohler2005a}.

The relative noise strength is characterized by the Fano factor $F =
S/e|I|$ which provides information about the transport mechanism
\cite{Blanter2000a}.  For a Poisson process it has value one,
while it becomes smaller for more regular transport.  In the present
context, we will speak of ``significant shot noise suppression'' if
the Fano factor lies clearly below 1/2.

\section{Time evolution}
In order to acquire insight into the system dynamics, we time
integrate master equation \eqref{ME} by a Runge-Kutta method.  We
start with an empty triple quantum dot just before the first double
pulse sets in and obtain the time evolution of the dot occupations
depicted in Fig.~\ref{fig:timeevolution}(b).  At the initial stage,
$t\lesssim0.1T$, dot~1 becomes populated by an electron that tunnels
in from the left lead.  Between the peaks of the double pulse [cf.\
panel~(a)], the electron essentially undergoes a direct transfer to
dot~3.  As non-adiabatic correction, dot~2 acquires some
occupation, which however stays clearly below the occupation of the
other dots.  After the double pulse, the electron tunnels further to
the right lead, and a next electron may enter at dot~1.  The fast
oscillation of $\rho_{22}$ when inter-dot tunneling is active can be
identified as tunnel oscillation between the middle dot and one of
its neighbors, because its frequency is roughly
$\Omega_\text{max}/2\pi$ \cite{Brandes2002a}.  For certain values of
the driving period $T$ [panel (c)], these oscillations become larger
and may interfere constructively, so that dot~2 acquires a larger
occupation.

If all tunnel matrix elements were constant in time, the population of
all three dots were about equal (not shown).  Thus, our transport
process must be dominated by CTAP if the occupation of the middle dot
is considerably smaller than the occupation of the other two dots.
In the ideal case, the steady state will be such that with each double
pulse, one electron is transported.  This gives rise to a regular
current, so that shot noise is suppressed.
In the following, we analyze this current and will find that the proper
course of the CTAP protocol correlates with small noise manifest in
the Fano factor.

\section{Steady state}
For computing the current noise, we need to solve simultaneously
master equation \eqref{ME} and the equation of motion for $R_1$, as
discussed above.  Since we are interested in the long-time behavior,
we eliminate transient effects by considering the results only after
all elements of the density matrix have changed by less than 0.25\%
with respect to the previous period.  Convergence is typically reached
after five to ten double pulses.

\begin{figure}
\centerline{\includegraphics{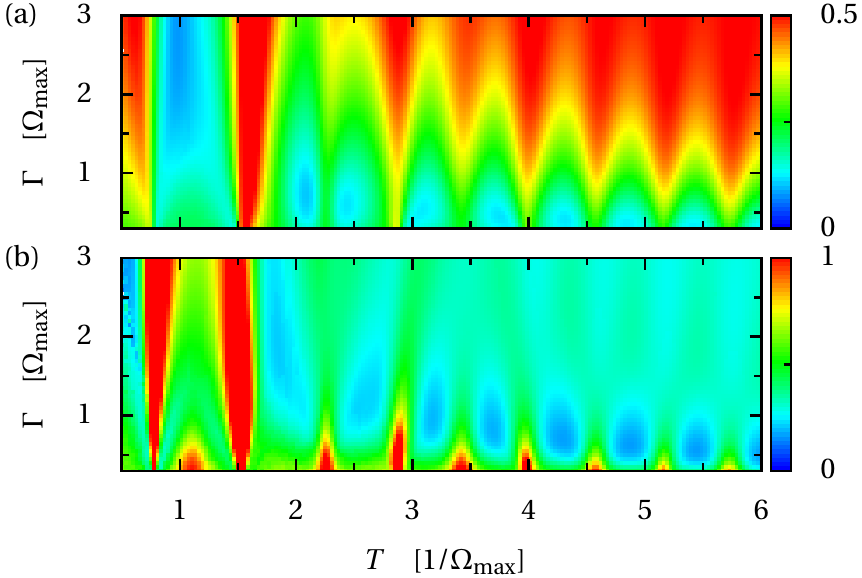}}
\caption{(Color online) (a) Time averaged population of dot~2,
$\bar\rho_{22}$, and (b) Fano factor $F=S/e|I|$ in the steady-state
limit as function the pulse repetition time $T$ and the dot-lead
coupling $\Gamma$.  For graphical reasons, the data are clipped at the
values 0.5 and 1, respectively.}
\label{fig:2d}
\end{figure}
Figure~\ref{fig:2d} provides an overview of the global behavior as
function of the driving period $T$ and the dot-lead coupling $\Gamma$.
For the time-averaged occupation of dot~2 shown in panel (a), we
distinguish three regions.  First, for $T\lesssim
1.5/\Omega_\text{max}$ in which the driving is too fast, such that
non-adiabatic effects play a strong role, and we cannot expect CTAP to
work properly.  Accordingly, the average population $\bar\rho_{22}$
may take any value.  The corresponding Fano factor [panel (b)] is
about 1/2 or larger, i.e., there is no relevant shot noise
suppression.  Notice the similarity of the dot occupation with
the one observed in Ref.~\cite{Greentree2004a}, despite that there
transient dynamics with phenomenological decoherence was considered.

For larger driving periods, we find a broad region in which the
population of the middle dot is rather low, unless the dot-lead
coupling $\Gamma$ becomes too large.  In particular, we find islands
in which both the average and the maximum of the mean occupation is as
small as $\bar\rho_{22} \approx\rho_{22} \approx 0.1$.  The plot for
the Fano factor exhibits a similar structure, where both $F$ and
$\bar\rho_{22}$ assume small values for roughly the same parameters.
This provides a first hint that our generalized CTAP gives rise to
regular, low-noise electron transport, in particular when $\Gamma$ is
of the order $\Omega_\text{max}$.  Interestingly, for certain values
of $T$, dot~2 becomes occupied and the Fano factor increases
accordingly.  This can be attributed to the mentioned tunnel
oscillations, because the distance between two peaks increases
linearly with the amplitude $\Omega_\text{max}$.
\begin{figure}
\centerline{\includegraphics{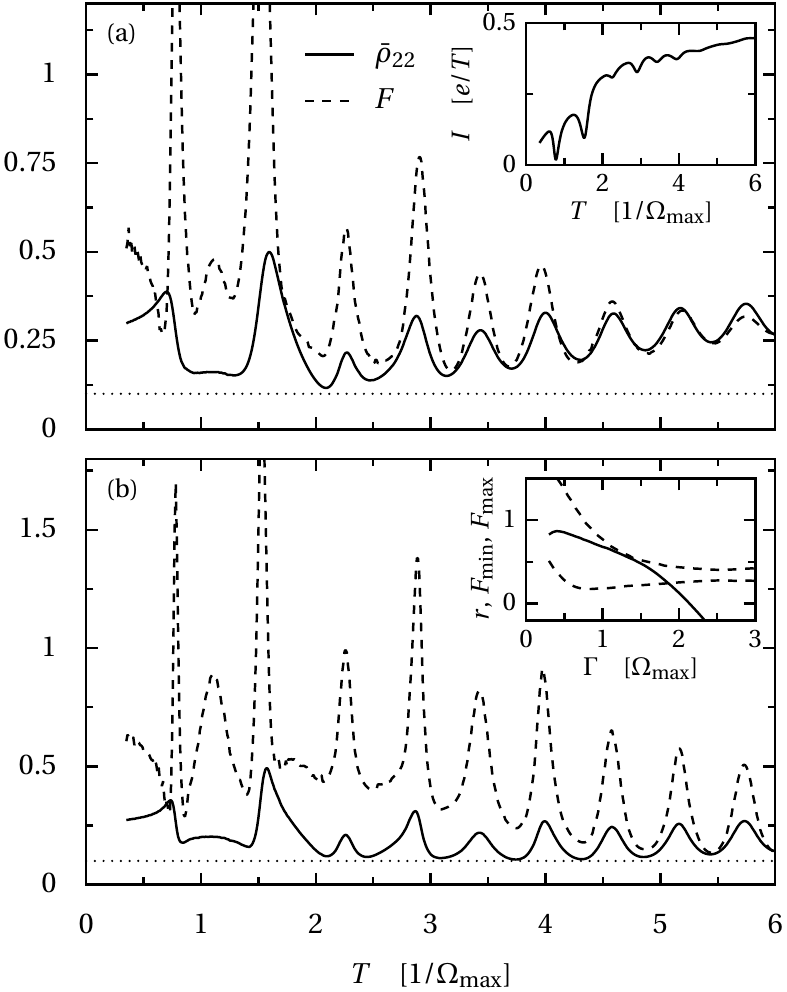}}
\caption{
Fano factor $F=S/e|I|$ and time-averaged population $\bar\rho_{22}$
for the dot-lead couplings (a) $\Gamma=\Omega_\text{max}$ and (b)
$\Gamma=0.5\Omega_\text{max}$.  The horizontal dotted line marks the
value $0.1$ which is the minimal reminiscent population of
dot~2.  Insets: (a) dc current corresponding to the main plot in units
of electron per pulse. (b) Correlation coefficient \eqref{corrcoeff}
of the Fano factor and~$\bar\rho_{22}$ in the range
$[2/\Omega_\text{max}, 4/\Omega_\text{max}]$ (solid line) and the
corresponding minimum and maximum of the Fano factor (dashed).
}
\label{fig:cut}
\end{figure}

For a more quantitative analysis, we depict in Fig.~\ref{fig:cut} the
time-averaged steady-state occupation $\bar\rho_{22}$ and the Fano
factor for two relatively small values of the dot-lead coupling
$\Gamma$.  This demonstrates that the population of dot~2 can be as
low as $0.1$, while it becomes roughly $1/3$ in the intermediate
region.  Accordingly, the Fano factor changes from a significantly
sub-Poissonian value to $F\approx 1$.  The dc current [inset of
Fig.~\ref{fig:2d}(a)], by contrast, exhibits only minor signatures of
CTAP.  This underlines that it is indeed the noise which provides
information about the nature of the transport process.  Moreover, it
reveals that for the relatively small $T$ considered, on average only
each second double pulse transports an electron.  This value can be
improved by increasing $T$ which, however, requires a smaller dot-lead
tunneling $\Gamma$.

We quantify the relation between the population $\bar\rho_{22}$ and
the Fano factor $F$ by the correlation coefficient
\begin{equation}
\label{corrcoeff}
r = \frac{\langle F,\bar\rho_{22}\rangle}{\Delta F\, \Delta\bar\rho_{22}}
\end{equation}
between these quantities for $T$ in the relevant interval
$[2/\Omega_\text{max}, 4/\Omega_\text{max}]$.  The numerator denotes
the covariance, while the denominator contains the standard
deviations, such that $r=1$ for ideal correlation.  The inset of
Fig.~\ref{fig:cut}(b) shows that for $\Gamma\lesssim
\Omega_\text{max}$, the correlation is rather high, typically $0.75$
or larger.  Moreover, the difference between the maximum and the
minimum of the Fano factor, $F_\text{max}-F_\text{min}$, is roughly
$0.5$, i.e., sufficiently large to be discriminated in an
experiment \cite{Barthold2006a}.  For larger values of $\Gamma$, the
Fano factor and the population become significantly anti-correlated.
There however, $F_\text{max}-F_\text{min}$ is quite small and may be
below the experimental resolution.  Therefore we do not further
discuss the regime $\Gamma\gtrsim1.5\Omega_\text{max}$.

In a possible experimental realization of steady-state CTAP, the
oscillatory dependence of the Fano factor as function of the pulse
distance $T$ may serve as indicator.  Most significant results are
expected for $\Gamma \lesssim \Omega_\text{max}$.  Thus, one may use a
triple quantum dot whose four barriers have similar transparency.
Negative gate voltages applied to the inner barriers allow one to
control the tunnel matrix elements.  Since the average current
consists of roughly half electron per double pulse, a current
$I\approx 10\,\mathrm{pA}$ requires a repetition rate of 100\,MHz
\cite{Hayashi2003a}.  The maximal tunnel rates should be considerably
larger than that, say, by a factor 10, which means that $\Gamma$ and
$\Omega_\text{max}$ should be of order $10\,\mathrm{\mu eV}$.

\section{Conclusions}
We have proposed a CTAP protocol for a triple quantum dot in transport
configuration, i.e., when the first dot and the last dot are coupled
to electron source and drain.  It induces a steady-state transport in
which electrons tunnel non-locally from the first to the last dot,
practically without populating the middle dot.  The noise properties
of the resulting current depend on the size of the reminiscent population.
The difference is manifested in the Fano factor which, thus, represents
a fingerprint of the proper course of the protocol.  Such indirect
evidence for CTAP is particularly useful, because any direct
measurement of the dot occupations would act back on the system and, thus,
unavoidably influence the effect to be substantiated.
A particular feature of steady-state CTAP is that it fails for certain
distances between pulse pairs.  Thus, upon changing this distance, one
obtains a sequence of sizable shot noise suppression and shot noise
enhancement, which should be measurable by standard techniques.
Thus, our proposal may initiate the experimental realization of this
intriguing non-local transport phenomenon.

This work was supported by the Spanish Ministry of Economy and
Competitiveness through Grant No.\ MAT2011-24331 and by the Marie Curie
Initial Training Network through Grant No.\ 234970.

%

\end{document}